\documentclass[twocolumn]{article}

\AtBeginDocument{%
  \providecommand\BibTeX{{%
    \normalfont B\kern-0.5em{\scshape i\kern-0.25em b}\kern-0.8em\TeX}}}

\usepackage{booktabs} 
\usepackage{graphicx}
\usepackage[utf8x]{inputenc}
\usepackage[T1]{fontenc}
\usepackage[textsize=footnotesize,linecolor=blue,backgroundcolor=yellow]{todonotes}
\usepackage{enumitem}
\usepackage{placeins}
\usepackage{algorithm}
\usepackage{algpseudocode}
\usepackage{hyperref}

\begin{document}
\title{A Revised Taxonomy of Steganography Embedding Patterns}

\author{Steffen Wendzel$^{1,2}$ \and Luca Caviglione$^3$ \and Wojciech Mazurczyk$^{4,2}$ \and Aleksandra Mileva$^5$ \and Jana Dittmann$^6$ \and Christian Krätzer$^6$ \and Kevin Lamshöft$^6$ \and Claus Vielhauer$^{6,7}$ \and Laura Hartmann$^{1,2}$ \and Jörg Keller$^2$ \and Tom Neubert$^{6,7}$}

\date{$^1$ Worms University of Applied Sciences, Worms, Germany\\
      $^2$ FernUniversität in Hagen, Hagen, Germany\\
      $^3$ National Research Council of Italy, Genova, Italy\\
      $^4$ Warsaw University of Technology, Warsaw, Poland\\
      $^5$ University Goce Delcev, Stip, North Macedona\\
      $^6$ University of Magdeburg, Magdeburg, Germany\\
      $^7$ Brandenburg University of Applied Sciences, Brandenburg, Germany\\~\\
      \small{\today}
    }

\maketitle

\subsubsection*{Abstract}
Steganography embraces several hiding techniques which spawn across multiple domains. However, the related terminology is not unified among the different domains, such as digital media steganography, text steganography, cyber-physical systems steganography, network steganography (network covert channels), local covert channels, and out-of-band covert channels. To cope with this, a prime attempt has been done in 2015, with the introduction of the so-called \textit{hiding patterns}, which allow to describe hiding techniques in a more abstract manner. Despite significant enhancements, the main limitation of such a taxonomy is that it only considers the case of network steganography.

Therefore, this paper reviews both the terminology and the taxonomy of hiding patterns as to make them more general. Specifically, hiding patterns are split into those that describe the \textit{embedding} and the \textit{representation} of hidden data within the cover object.

As a first research action, we focus on embedding hiding patterns and we show how they can be applied to multiple domains of steganography instead of being limited to the network scenario. Additionally, we exemplify representation patterns using network steganography. Our pattern collection is available under \url{https://patterns.ztt.hs-worms.de}.

~

\textbf{Keywords:} Network Steganography, Covert Channels, Terminology, Taxonomy, Information Hiding, Science of Security, Information Security, Patterns, PLML, Cyber Security.

~

\textbf{Published:} In Proc.\ of the 16th International Conference on Availability, Reliability and Security (ARES'21), August 17--20, 2021, Vienna, Austria.

~

\textbf{DOI Link:} \url{https://doi.org/10.1145/3465481.3470069}

~

\section{Introduction}
\textit{Steganography} is the art and science of hiding information in so-called cover objects, e.g., a secret message is embedded inside a digital file, network packet, or written text. Its counterpart, \textit{steganalysis}, aims at detecting, preventing, and limiting steganography. Several attempts have been made to define the fundamental terminology and its domains, such as text steganography, digital media steganography, or network steganography \cite{infterm,petitcolas1999information,Fridich2009,NIHbook}.
One of these attempts to unify and refine the terminology led to the systematization of steganographic techniques in precise, general, and abstract templates, defined as \textit{hiding patterns} \cite{CSURpaper}.
Each hiding pattern is described via the \textit{Pattern Language Markup Language} (PLML) allowing to outline all the various templates in a unified manner. By using PLML, patterns can be derived from each other forming a taxonomy, and they can also be linked or composed.  

Despite being progressively adopted by the scientific community (140+ citations as of June 2021), hiding patterns have some limitations. First, hiding patterns are only defined for the sub-discipline of network steganography. 
Second, network-specific hiding patterns cannot be directly applied to other domains of steganography. Thus, the absence of unified terminology and taxonomy as well as the impossibility of exploiting overlaps to generalize core concepts, are key issues preventing the adoption of the pattern-based paradigm by a wide audience. At the same time, a precise terminology in the ever-growing research domain of steganography is a real need, as to limit scientific re-inventions and terminological inconsistencies \cite{wendzel2015creativity}. For instance, distinguishing between the sender and receiver side of various patterns as proposed in \cite{ARES2018Patterns} is not an optimal solution and it could lead to ambiguities. 

Therefore, in this paper we aim at addressing the aforementioned issues. We first summarize the characteristics of three steganography domains, i.e., network steganography, digital media steganography, and text steganography. We then present a methodology to unify the description of hiding patterns in a domain-overlapping manner. Compared to previous works (see \cite{CSURpaper, ARES2018Patterns}) emphasis will be put to provide a less ambiguous distinction between the \textit{embedding} process and the \textit{representation} of hidden information within the cover object. We especially focus on the embedding patterns for which a novel taxonomy is provided while existing patterns are integrated into a list of representation patterns. 

The rest of this paper is structured as follows. Sect.\ \ref{Sect:ExistSubDomains} explains the characteristics of three key domains of steganography, namely network steganography, digital media steganography and text steganography. It further points out the limits of the existing network steganography-based taxonomy. Sect.\ \ref{Sect:Methods} explains our methodology while, Sect.\ \ref{Sect:NewTaxonomy} presents our unified terminology and taxonomy of embedding patterns and exemplifies representation patterns using network steganography. Sect.\ \ref{Sect:AnticipatedDevs} highlights the anticipated future developments of steganography that might influence our pattern-based taxonomy and, finally, Sect.\ ~\ref{Sect:ConclFut} concludes the paper and provides an outlook on future work.

\section{Analysis of Existing Steganography Domains}\label{Sect:ExistSubDomains}

\begin{figure}[tb]
\centering
\includegraphics[width=0.98\linewidth]{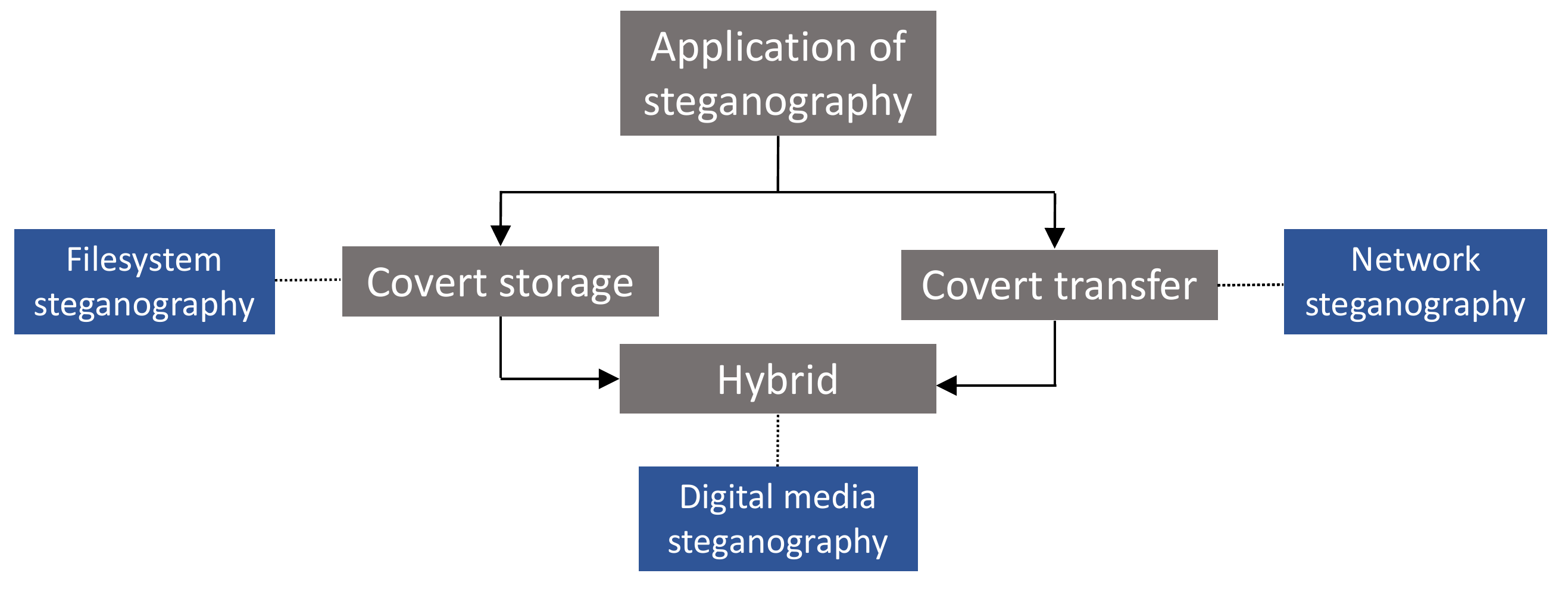}
\caption{Various applications of steganography.}\label{fig:steg-app}
\end{figure}

In general, steganographic techniques can be utilized either to enable a \textit{covert transfer} or \textit{covert storage}, as well as in a combined manner as depicted in Fig.\ \ref{fig:steg-app}. In both cases, the secret information is \textit{embedded} into a \textit{cover object}, which should be selected to not represent an anomaly and have a suitable embedding capacity. Typically, a steganographic application or technique is closely related to the features characterizing the chosen hidden data carrier. In more detail, for the case of covert transfer, a \textit{covert sender} (CS) transmits secret information to a \textit{covert receiver} (CR). Even if many mechanisms for covert transfer exist, the most popular group of information hiding solutions exploit network traffic and protocols \cite{zander}. Instead, for the case of covert storage, the steganographer is interested only in storing sensitive data on a local information carrier (e.g., on a hard drive), in such a way that the data cannot be spotted by a third party observer unaware of the information concealment. An example of such a technique is filesystem steganography where some additional overlay filesystem for data hiding purposes is created by using features like the unused space in partially-allocated blocks \cite{Khan:2011}.

Finally, for some cover objects, it is possible to perform covert transfer or covert storage depending on the required application. This is the case, for instance, of digital media steganography where one can perform a hidden data exchange by embedding secret data into the content transferred by services like video or audio streaming, or even if one sends an email with an image containing secret information. Alternatively, the steganographer can utilize digital images as a vault to locally store his/her secrets \cite{stegsur}. Recently, another new set of techniques emerged that combines the covert transfer (over network covert channels) and the covert storage (within the caches of network protocols), which is called a \textit{Dead Drop} \cite{Schmidbauer:DeadDrops:ARP}.

The remainder of this section highlights how major steganography domains differ in terms of their cover objects and embedding strategies.

\paragraph{Network Steganography}
As hinted, the principal characteristics of network steganography are already covered by the existing terminology (see e.g., \cite{CUING:CACM} and the references therein). In essence, the main cover objects used in network steganography are provided by manipulating or injecting information in some digital artifacts belonging to the network traffic, e.g., the header or the payload of a Protocol Data Unit (PDU) as well as in the behaviors of flows/conversations consisting of a coherent sequence of packets. In general, two main flavors of network steganography exist: \textit{i}) direct embedding of data within the PDU, or \textit{ii}) by modulating the timing or the sequence of adjacent/succeeding packets. Compared to other steganography domains, the goal of network steganography is not to \textit{store} but to \textit{transfer} the data \cite{CUING:CACM}. The capacity of a steganographic method targeting network is limited by the traffic type and the length of a transmission. Typically, this leads to a slower embedding process compared to digital media steganography \cite{zander, stegsur}. The data is hidden in an ephemeral manner and the application of network steganography can increase delays and packet loss. This can impact on the stealthiness of the resulting covert transmission due to the reduction of some functionality provided by the protocol or a degradation of the transmission quality \cite{CUING:CACM}.

\paragraph{Digital Media Steganography}
The term digital media steganography (or short: media steganography) addresses the wide field of digital steganography research and development focusing on digital media (i.e., media encoded in machine-readable formats) as cover data for a plausible, secured and hidden communication. Digital media were initially designed to address the human audio-visual system (by delivering information to a screen and/or loudspeaker) and include many heterogeneous forms as images, audio data, videos, 3D models, etc. As with the media themselves, digital media steganography comes in a wide variety of different types that can be classified by various  categories. In particular, digital media steganography can focus on the media type(s) (e.g., audio steganography), the transmission method (e.g., data as spatial image or as audio stream vs.\ audio files) and the basic strategy concerning the existence and plausibility of a cover data (such as a data stream to embed into).
Three paradigms for the message embedding can be applied: steganography by modification, by synthesis and by selection.
Further in the case of steganography by modification, the basic coding strategies of message insertion (i.e., where to embed in the cover data), the structure of how to embed the message in the cover data (usually represented as a signal or coded signal data), as well as the usage of the steganographic key are common categories.

Since becoming an active research field in the 1990s, a great number and wide variety of scientific works have been published on media steganography and steganalysis. The vast majority of these publications (as well as most of the tools available) have been focusing on image steganography as the most prominent sub-domain in this field \cite{Fridich2009}.

It can be stated that any continuous digital media (in the sense of temporally-changing media content) can be designed both for covert storage and covert transfer. This obviously applies mainly to audio and video, which can be streamed or stored as files. Recently, streaming services received an increasing degree of interest, as they appear to become the new main form of media delivery and consumption in entertainment.

The capacity of digital media steganography is limited by the type and size of the digital media. For digital media steganography, capacity always depends on two other characteristics to be achieved: robustness and imperceptibility for the detection of the hidden message (also related to undetectability).
Some methods of media steganography can survive conversion to another format, but a plausible cover object is always required. The application of digital media methods might decrease the quality of the cover object (e.g., image quality).

\paragraph{Text Steganography}
This distinct branch of steganography relies on hiding information in textual messages and textual documents as cover data, including those in magazines, newspapers, word processing documents, personal notes, and music notes -- just to mention a few. In contrast to digital media steganography, it uses manipulation of some lexical, syntactic or semantic features of the text content, modification of different features of the text’s elements (e.g., characters, paragraphs, sentences, words, lines) or generation of a new text that simulate some features of the normal text. Several examples of such techniques are presented in \cite{petitcolas1999information} and more recently in \cite{GuptaGupta2011}.
The latter has identified the following concepts as embedding principles in the literature: \textit{i)} word spelling, \textit{ii)} semantic method, \textit{iii)} line shifting, \textit{iv)} abbreviation, \textit{v)} word shifting, \textit{vi)} syntactic method, and \textit{vii)} new synonym text. Since at least three of these (i.e., \textit{ii}, \textit{iv}, and \textit{vii}) can be considered of purely semantic nature, and since in comparison to digital media steganography, text steganography also involves printed (non-digital) text, the distinction between them and the field of digital media steganography seems reasonable.

Similar to digital media steganography, text steganography allows the permanent hiding of information as the texts are not of ephemeral nature like network traffic. To this end, the vast majority of proposed concepts can be categorized as covert storage techniques. However, concepts of embedding hidden information in text streams (e.g., keystrokes or scrolling text) appear feasible.

The capacity of text-based steganographic methods is mainly limited by the size and structure (including grammar, sections and use of white-spacing) of a text. However, a suitable cover text is required to make it plausible as auto-generated texts might appear synthetic to an observer. Similar to digital media steganography, text steganography may decrease the quality of the cover object, even if imperceptible.

\paragraph{Other Steganography Domains}
Additional domains of steganography bring different characteristics with them. For instance, in filesystem steganography, the cover object might be a file, unused space in a partially allocated block, cluster distribution of an existing file \cite{Khan:2011}, or an inode \cite{Eckstein:2005}. In cyber-physical systems (CPS) steganography, a value might be embedded into a sensor value \cite{Ulz:2019}, an actuator state or unused registers \cite{IoTStego17}, or into the control logic of a PLC \cite{Krishnamurthy:2018}. Hidden data might even be embedded into the number of cyber-physical events of some machine. Hildebrandt et al.\ published the only available pattern-based classification for CPS steganography \cite{hildebrandt2020threat}, built on top of the existing one for network steganography. However, their taxonomy adds additional categories, namely for firmware accessible and program accessible patterns. 

\paragraph{Summary}
When we look at the aforementioned steganography domains, it becomes clear that cover objects appear to be highly different, involving events, values or states, not just files or packets. 
For this reason, the novel taxonomy must allow for the inclusion of highly heterogeneous events, based on a taxonomy that incorporates events, values, and states to unify the patterns of steganography.

In general, a unified theory/taxonomy can be more suitable for a research area than multiple domain-specific theories in a similar manner that universal programming languages can be advantageous over domain-specific programming languages (see sect.~\ref{Sect:Limitations} for the limitations of the domain-specific approach).

\subsection{Limitations of the Current Approach}\label{Sect:Limitations}

While there are several advantages of unifying the terminology and taxonomy of hiding methods (e.g., they help structuring steganalysis processes), there are also certain limitations with the current pattern-based taxonomy which shall be addressed by our work:

\begin{enumerate}
    \item Currently, the available terminology and taxonomy of hiding patterns are limited to network communications, neglecting other domains of steganography.
    \item The level of abstraction of the current taxonomy does not allow for the inclusion of non-network patterns. For instance, \textit{user-data} from the perspective of network steganography might be a digital media payload. However, from a digital media perspective, the network steganography context would not matter. Thus, current pattern names, e.g., \textbf{Payload Field Size Modulation}\footnote{In this paper, pattern names are written in \textbf{bold} font.}, and taxonomy, e.g., \textit{user-data awareness}, are not fully suitable. A novel taxonomy should therefore discard \textit{domain-specific} abstractions. For instance, a least significant bit(s) (LSB) method applied to an image file and an LSB method applied to a network packet share the same concept and it is the concept that matters.
    \item The current set of available hiding patterns does not discriminate between the \textit{embedding} process and the \textit{representation} of hidden information in a carrier, rendering the interpretation of existing hiding patterns ambiguous.
    \item Some of the original patterns are actually hybrid patterns that should be broken down into their atomic pieces to describe them clearly (see, e.g., the \textbf{Sequence Modulation} pattern in Sect.\ \ref{hybrid_patterns} (\ref{Ref:SequenceModulation})).
    \item The current systematic categorization of patterns partially follows the ``open science'' paradigm by providing information about new patterns through a freely accessible website. However, the inclusion of additional scientists and research groups was not actively sought, which we aim to change by encouraging scientists to participate in our consortium.
\end{enumerate}

\section{Methodology}\label{Sect:Methods}

We set up a consortium consisting of eleven experts from seven institutions located in four countries. During regular consortium meetings, the following methodology emerged.
Given the success and the functionality of hiding patterns, we decided to keep the concept of patterns for the new taxonomy. It was further agreed that the consortium will stick to the PLML-based pattern specification that was already applied by \cite{CSURpaper}. PLML provides a comparable and unified systematic for the description and management of patterns \cite{PLML11} that is also applied in other areas, such as software engineering. A PLML-based description contains certain attributes, such as a name for the pattern, aliases, an illustration, code snippets, evidence in form of references, example cases, and links to related patterns \cite{PLML11} --- just to mention a few. A PLML-based specification also allows to exploit existing methodology, such as the unified description method for hiding techniques \cite{JUCS:UnifiedDescriptionMethod} and the existing framework for determining whether some hiding technique represents a new pattern, or not \cite{wendzel2015creativity}. 
Furthermore, PLML enables easy indexing,
extensibility and linkage of patterns to keep the provided taxonomy up-to-date on the long run. By allowing the inclusion of \textit{aliases} in PLML-based specifications, different terminology can be unified in a common term as well, limiting the chance for so-called scientific re-inventions \cite{wendzel2015creativity}.

\section{A Novel Taxonomy of Hiding Patterns}\label{Sect:NewTaxonomy}

This section presents our taxonomy for hiding patterns in a way that incorporates the characteristics of the discussed steganography domains. 
The central aspect of our taxonomy is to split all patterns into two categories:

\begin{enumerate}
    \item \textit{Embedding Patterns} describe how secret information is embedded into a cover object, such as an image file or a network packet.
    \item \textit{Representation Patterns} describe how the secret information is represented in a cover object.
\end{enumerate}

It must be noted that when secret data is embedded via the pattern A, it is not necessarily represented by the same pattern, but it can be. Two examples illustrate this statement:

\begin{enumerate}
    \item \textit{Embedding Pattern $=$ Representation Pattern:} CS sends an IP packet to CR in which it manipulates the least significant bit of the Time to Live (TTL) field. CR reads the very same value. Thus, the embedding uses the \textbf{State/Value Modulation} pattern while the hidden information is also represented by this pattern.
    \item \textit{Embedding Pattern $\neq$ Representation Pattern:\label{EmbedNeqRepPattern}} Let us assume an indirect covert channel, where the CS exploits functionality of a central element that is observed by the CR. Let us further assume that a third-party client is getting disconnected from the central network node if some specific \textit{value} is sent to it. The CS would then use the so-called \textbf{Value Modulation} pattern to cause a disconnect of a certain client from the central element. However, the CR might only be able to poll the list of (re-)connected clients at the central element, i.e., the hidden information would be represented by the \textbf{Artificial Reconnections} pattern introduced in \cite{COSE:MQTT5}.
\end{enumerate}

\subsection{Justification of Taxonomy Design Decisions}

In previous works \cite{CSURpaper,NIHbook,ARES2018Patterns}, several taxonomy layers specific to network steganography have been proposed, which we modified or even discarded for the new taxonomy due to reasons given in the following subsections.

\subsubsection{Previous Terminology Was Based on Packets and Messages}
It arose early during discussions that the current network steganography hiding patterns terminology does not fully reflect other steganography domains. For instance, the pattern \textbf{Inter-packet Times} relates only to network packets and a more generic pattern should thus be named \textbf{Event/Element Interval Modulation}. A similar case is the \textbf{Message Timing} pattern, which has been renamed to \textbf{Event Occurrence}. Similarly, \textbf{Value Modulation}, \textbf{Message Timing} and other patterns need to reflect non-network specific aspects, such as states of cyber-physical systems, texts and filesystems, which resulted in novel terms, such as \textbf{\emph{State}/Value Modulation}.

\subsubsection{Previous Terminology Focused on Payload}\label{Sect:UserdataPayload}
Another issue when transferring the network steganography terminology to the broader steganography context was the term \textit{payload} as there was a set of payload-specific patterns. From a network perspective, an image nested in a packet would be the payload, but the image would be the major focus in digital media steganography, where the network packet headers would be irrelevant. Thus, we decided to discard the term payload as well as the taxonomy abstraction between payload and non-payload. We further removed the terms \textit{user-data} (as it referred to payload) and the linked terms \textit{user-data aware} and \textit{user-data agnostic}.

\subsubsection{Syntax vs.\ Semantics}
We decided not to discriminate between patterns that modify (corrupt) the syntax and those that modify the semantics of a cover element. This is rooted in the fact that several patterns can modify both. For example, let us assume that we apply our new pattern \textbf{Elements/Features Positioning}, which modulates the position of an \textit{element} (we simply use a \textit{word} as an element) in the sentence \texttt{Joe has the right \textbf{not} to sign the document after 10.00 o'clock}. So, Joe would be allowed to reject signing the document after 10.00.
When we shift the position of the word ``not'' we can either break the original meaning of the sentence (\texttt{Joe has \textbf{not} the right to sign the document after 10.00 o'clock}, i.e., now Joe is not allowed anymore to sign the document after 10.00, even if he would like to do so) or the grammar (syntax) (\texttt{Joe has the right to sign the \textbf{not} document after 10.00 o'clock}). Similarly, a structured network packet header could be used to exemplify this aspect, cf.\ \cite{WendzelKeller:CMS12}.

\textit{Structure-preserving:} In this context, we consequently decided to discard the distinction between structure-preserving and structure-modifying non-temporal methods.

\subsubsection{Temporal vs.\ Non-temporal Patterns}\label{TemporalVSNontemporal}
While temporal hiding patterns are considered those that modulate timing behavior (e.g., timing between succeeding network packets), non-temporal hiding patterns are those that do not modify temporal aspects, at all. However, non-temporal patterns can be applied in a \textit{sequence}, though. For instance, if the \textbf{Elements/Features Positioning} pattern is applied to one IPv4 packet header and places some IP option at a specific position in the \textit{list of options}, this is a non-temporal pattern: the sequence of bits is not considered temporal and the packet is sent in one piece. However, if the \textbf{Elements/Features Positioning} pattern is applied to several succeeding IP packets in a row, the pattern is still considered as non-temporal. Its succeeding application \textit{might} result in transmissions errors if one packet overruns another, due to temporal behavior, but the \textit{embedding process was not directly focusing or considering this temporal behavior}, nor would the data be \textit{represented} by the temporal behavior (but instead by the order).

\textit{Discarding Protocol-awareness of Temporal Patterns:} To ease the accessibility of our taxonomy, we discarded the previous differentiation between \textit{protocol-aware} and \textit{protocol-agnostic} temporal patterns. Communication protocols are not the core subject of the new taxonomy anymore. Moreover, methods can be protocol-\textit{aware} at one layer and protocol-\textit{agnostic} at another. For instance, the original \textbf{Inter-packet Times} (\cite{CSURpaper}) pattern requires at least awareness of low-level frames but it does not need awareness of higher-layer protocols encapsulated into the frames. Additionally, if the \textbf{Inter-packet Times} pattern would operate on a higher level, it would require the understanding of frame structures, packet structures etc., e.g., when timings of UDP datagrams are modulated, the IPv4/IPv6 structure must be known.

\subsubsection{Discarding ICS-specific Taxonomy Categories}
The categorization between \textit{firmware accessible} and \textit{program accessible} patterns as proposed by Hildebrandt et al.\ \cite{hildebrandt2020threat} was dropped for the same reasons as network-specific categorizations: they do not fit into all domains. ICS-specific patterns will be addressed in follow-up works.

\subsubsection{Extendability of the Taxonomy}
A key criteria for the design of our taxonomy is its extendability. As mentioned in Sect.~\ref{Sect:Methods}, PLML will be used as a tool to achieve extendability. With PLML, patterns can be updated (also on the website) to reflect changes; they can also be added if new patterns are discovered and aliases as well as relations between patterns can be updated.

\subsection{Naming Conventions}
Hiding patterns are identified by a number (Sect.~\ref{Sect:Enumeration}) and a name (Sect~\ref{Sect:PatternNames}).

\subsubsection{Enumeration of Patterns}\label{Sect:Enumeration}
As embedding patterns are of a generic nature, they are not required to reflect any steganography domain in their enumeration. Their enumeration follows the convention \textbf{E[TN]n}, where \textbf{[TN]} means that either \textbf{T} or \textbf{N} are used. Temporal embedding patterns follow the enumeration convention \textbf{ETn} (\textit{embedding; temporal, number $n$}) while non-temporal patterns follow the enumeration convention \textbf{ENn} (\textit{embedding; non-temporal, number $n$}). Sub-patterns add an additional number followed by a dot, e.g., \textbf{ETn.x} (the $x$-th sub-pattern of the temporal embedding pattern ET$n$). Additional hierarchy layers can be represented accordingly, such as \textbf{ETn.x.y} or even \textbf{ETn.x.y.z}, if necessary.

Representation patterns are always domain-specific and follow the enumeration convention \textbf{R[TN]nD}, where \textbf{R} tells us that it is a representation pattern and \textbf{T} and \textbf{N} differentiate between temporal and non-temporal hiding patterns (same as above). \textbf{n} is again the number of the hiding pattern. The only novelty is the parameter \textbf{D=[ndtcf]}, which represents the steganography domain, of which the following are defined so far: \textbf{n} (network steganography), \textbf{d} (digital media steganography), and \textbf{t} (text steganography). We additionally define (but not use in this paper) the steganography domains \textbf{c} (cyber-physical steganography) and \textbf{f} (filesystem steganography). This convention might be extended in the future to reflect additional steganography domains. For instance, the representation pattern \textbf{RT1t} tells us that it is a temporal representation pattern with the number 1 and it belongs to text steganography.

\subsubsection{Naming of Patterns}\label{Sect:PatternNames}
The naming of patterns follows a clear structure. A pattern name contains three components. First, its number, second, the \textit{modifiable object} (e.g., \emph{Event} or \emph{Feature}) and, third, the \textit{action} of a pattern (e.g., \emph{Modulation} or \emph{Occurrence}).\footnote{Please note that the previously introduced term \textit{cover object} is not meant when we refer to a \textit{modifiable object}.} The full pattern name separates all three components by a space, e.g., \textbf{ET2.\ Event Occurrence}. Sect.~\ref{sect:Glossary} provides a list of objects and actions. However, additional objects and actions might be defined in future work.

\subsection{Glossary}\label{sect:Glossary}
As a preliminary, we introduce some basic terminology, which will be used in the remainder of the paper. Even if the creation of a non-ambiguous vocabulary for steganographic applications is outside the scope of this work, reducing possible confusions or overloading of terms is fundamental to not void the efficiency and expressiveness of the taxonomy. Specifically, the term \textit{modifiable object} we define as the \textit{general} object type that will be used to contain the secret information. The process of hiding data within the cover depends on the used mechanism or pattern. In the following, we refer to such a process as \textit{embedding}, \textit{injecting} or \textit{hiding}. The term \textit{modulating} will be used in case of ambiguities, especially to highlight that the secret information is not directly stored but encoded by means of variations of the cover object. The amount of data that can be hidden will be denoted as the \textit{capacity}. 

In general, patterns can be used both to describe the process of hiding information for storage purposes as well as to secretly move data among two endpoints. To avoid burdening the text, when the ``transmissional'' nature of the embedding process is not obvious, we will explicitly identify the covert sender and receiving side as to emphasize the origin and the destination of the steganographic communication.

\begin{table*}[t]
  \scriptsize
  \caption{Differentiation between the types of \textit{objects} used in this paper.}
  \label{tab:objects}
  \begin{tabular}{p{2.8cm}|p{2.5cm}p{2.8cm}p{1.8cm}p{2.6cm}p{2.2cm}}
    \toprule
    \textbf{Domain} & \textbf{Interval} & \textbf{Event} & \textbf{Element} & \textbf{Feature} & \textbf{State/Value} \\
    \midrule
    \textbf{network steganography} & time between packets & presence of flow; disconnect & network packet & size of packet; field of packet & value of header field; number of packets \\
    \hline \\
    \textbf{text steganography} & time between text notes sent & occurrence of character sequence & character & color of character & number of characters \\\\
    \hline \\
    \textbf{digital media steganography} & duration of audio file & occurrence of pre-defined sound in MP3 file & pixel of image & color of pixel & value of pixel; number of pixels in image \\
  \bottomrule
\end{tabular}
\end{table*}


For the specific case of defining the taxonomy as well as to describe patterns, the following formal definitions have been introduced: 
\begin{enumerate}
    \item Modifiable Objects (see, Tab.~\ref{tab:objects}):
    \begin{itemize}
        \item An \emph{Event} describes a (timed or forced) appearance, which can be composed of several elements, e.g., 1) the appearance of a predefined character sequence; 2) a predefined specific sound in a video; 3) network connection establishment, reset or disconnection.
        \item An \emph{Element} represents a single unit of a whole sequence, e.g., 1) a word/character of a text; 2) a pixel of an image; 3) a network packet of the whole flow.
        \item A \emph{Feature} characterizes a property of an element to be modulated, e.g., 1) the color of a character; 2) the attribute of a tag in vector graphics; 3) the field / the size of a network packet.
        \item An \emph{Interval} specifies the temporal gap between two events, e.g., 1) the duration of an audio file; 2) the time between sending a message and receiving the related acknowledgement.
        \item A \emph{State/Value} denotes a non-temporal numerical or positional quantity of an element, feature, or event, e.g., 1) the values of TCP header fields (feature value); 2) the x-y-z coordinates of a player in a 3D game.
    \end{itemize}
    \item Actions:
    \begin{itemize}
        \item An \emph{Occurrence} is the temporal location of a given element, feature, or event observed in the cover.
        \item A \emph{Modulation} of an element's (or event's) value (or state) is the selection of one particular value/state (out of multiple possible values/states).
        \item A \emph{Corruption} refers to the blind overwriting of an element, feature or state/value. 
        \item \emph{Enumeration} means that the overall number of appearances of something is altered.
        \item \emph{Repeating} refers to duplicating elements, events or features (multiple times). It can be considered a sub-form of the \textit{enumeration} action.
        \item \emph{Positioning} selects the non-temporal position of an element in a sequence of elements.
    \end{itemize}
\end{enumerate}

\subsection{Embedding Patterns}\label{Sect:NewTaxonomyEmbedding}

\begin{figure*}[th]
    \centering
\includegraphics[width=0.79\textwidth]{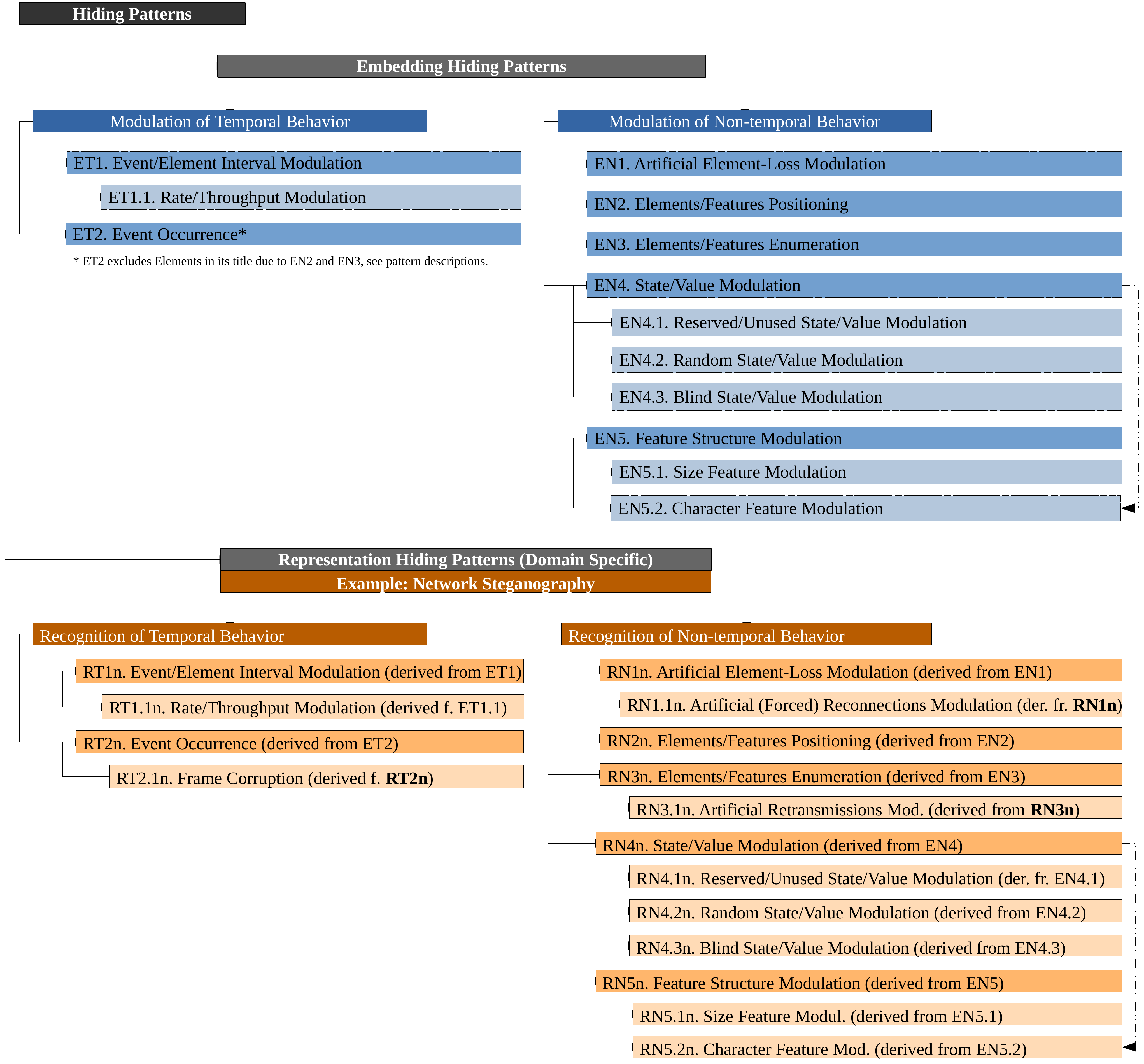}
\caption{The novel, general-purpose taxonomy of embedding hiding patterns for steganography (and exemplary representation hiding patterns for the network steganography domain).}\label{fig:taxonomy}
\end{figure*}

Our novel taxonomy of hiding patterns contains two major branches (see Fig.~\ref{fig:taxonomy}): patterns that describe how information is \textit{embedded} in a cover object and patterns that describe how embedded secret data is \textit{represented} in it.

\subsubsection{Modulation of Temporal Behavior}

The covert message is embedded by modulating how a behavior evolves in time.

\paragraph{ET1.\ Event/Element Interval Modulation}
The covert message is embedded by modulating the gaps between succeeding events/elements, for instance by: 1) modulating the inter-packet gap between succeeding network packets (elements) or between connection establishments (events); 2) modulating the time-gap between succeeding cyber-physical actions, such as acoustic beeps.
      \begin{enumerate}
          \item \textit{Rate/Throughput}: The covert message is embedded by alternating the rate of events/elements (by introducing delays or by decreasing delays). Here, several inter-event/element intervals have to be modified in a row to embed a secret message, i.e., the message is not embedded into particular inter-event/element timings but in the \textit{overall} rate/throughput. Examples: 1) modulating the packet rate while sending traffic to some destination (by decreasing/increasing delays between \texttt{send()} actions); 2) modulating the number of produced items per hour in a production facility.
    \end{enumerate}

\paragraph{ET2.\ Event Occurrence}
The covert message is encoded in the temporal location of events (in comparison to ET1.1, the \textit{rate} of events is not directly modulated but events are triggered at specific moments in time, moreover, ET2 can be a single event while ET1.1 needs a sequence of elements), e.g., 1) sending a specific network packet at 6pm; 
2) influencing the time at which a drone starts its journey to some destination (or its arrival time); 3) performing a disconnect at a certain time.
\\\emph{Note:} We did not include \textit{elements} into this pattern in favor of EN2 and EN3. See also Sections \ref{TemporalVSNontemporal}.

\subsubsection{Modulation of Non-temporal Behavior}

\paragraph{EN1.\ Artificial Element-Loss Modulation}
The covert message is embedded by modulating the artificial loss of elements. Examples: 1) dropping TCP segments with an even sequence number; 2) removing commas in sentences \cite{Bender:1996}.

\paragraph{EN2.\ Elements/Features Positioning}
The covert message is embedded by modulating the position of a predefined (set of) element(s)/feature(s) in a sequence of elements/features. Examples: 1) position of an IPv4 option in the list of options; 2) placing a drink on a table to signal a Go player to play more defensive; 3) placing a specific character in a paragraph.

\paragraph{EN3.\ Elements/Features Enumeration}
The covert message is embedded by altering the overall number of appearances of elements or features in a sequence. Examples: 1) fragmenting a network packet into either $n$ or $m$ ($n\neq m$) fragments; 2) modulating the number of people wearing a t-shirt in a specific color in an image file; 3) repeating an element/feature by duplicating a white space character (or not) in a text \cite{Bender:1996}.

\paragraph{EN4.\ State/Value Modulation}
The covert message is embedded by modulating the states or values of features, e.g., 1) performing intense computation to influence some temperature/clock-skew \cite{NIHbook}; 2) modulating other physical states, such as proximity, visibility, force, height, acceleration, speed, etc.\ of certain devices;
3) changing values of the network packet header fields (e.g., target IP address of ARP \cite{5541791}, Hop Count value in IPv6 \cite{lucena2005covert} or the LSB in the IPv4 TTL);
4) modulate the x-y-z coordinates of a player in a 3D multiplayer online game \cite{zander2008GamingCCs}.

    \begin{enumerate}
        \item \textit{Reserved/Unused State/Value Modulation}: The covert message is embedded by modulating reserved/unused states/values, e.g., 1) overwriting the IPv4 reserved field \cite{HandelSandford96}; 2) modulation of unused registers in embedded CPS equipment \cite{IoTStego17}.

        \item \textit{Random Modulation}: A (pseudo-)random value or state is replaced with a secret message (that is also following a pseudo-random appearance), e.g., 1) replacing the pseudo-random content of a network header field with encrypted covert content; 2) encoding a secret message in the randomized selection of a starting player in an online chess game.
        
        \item \textit{Blind State/Value Modulation}: Blind corruption of data, e.g., 1) blindly overwriting a checksum of a PDU to corrupt a packet (or not) to embed hidden information; 2) blindly overwriting content of a file in a filesystem, neglecting its file header; 3) blindly overwriting a TCP payload.
    \end{enumerate}

\paragraph{EN5.\ Feature Structure Modulation}
This hiding pattern comprises all hiding techniques that modulate the structural properties of a feature (but not states/values (EN4), positions (EN2) or number of appearances (EN3)). Examples include: 1) increasing/decreasing the size of succeeding network packets; 2) changing the color/style of characters in texts.

    \begin{enumerate}
        \item \textit{Size Modulation}: The covert message is embedded by modulating the size of an element, e.g., 1) create additional (unused) space in network packets for embedding hidden data, such as adding an ``unused'' IPv6 destination option \cite{IPv6DestOptCC}; 2) alternate the size of PNG files.

       \item \textit{Character Feature Modulation}: Modulation of different features in characters, such as color, size (scale), font, position or size of different parts in some letters, e.g., 1) using upper/lower case letters in HTTP or SMTP requests \cite{Dyatlov:2003}; 2) modulating the color of characters in text steganography.\\
       \textit{Relations:} Utilizes partially the same methods as EN4.\ State/Value Modulation (e.g., a HTTP header field's \textit{character} is also a \textit{value}). Thus, both are linked in see Fig.~\ref{fig:taxonomy}.
    \end{enumerate}

\subsubsection{Hybrid Embedding Patterns}
It must be noted that the hybrid application of embedding methods is feasible, too. For example, the LACK method for IP telephony uses ET1 (by applying artificial delays) and EN4 (by changing the value in the payload field) \cite{LACK}. As we exemplify in Sect.~\ref{Sect:Network_StegoRep}, hybrid \emph{representation} patterns exists as well.

\subsubsection{Example 1: Network Steganography}
As discussed, network steganography is a steganography domain for which hiding patterns were already defined. Thus, our embedding patterns were designed on the basis of the hiding patterns designed for network steganography as introduced by \cite{CSURpaper} and extended/updated by \cite{NIHbook,ARES2018Patterns,COSE:MQTT5}. For this reason, the embedding patterns match the known embedding strategies for network steganography.

Instead of separating patterns into timing and storage patterns, we favored the differentiation between temporal and non-temporal behavior, which is only loosely related to the original distinction. We verified that all previously known network steganography hiding patterns' embedding functionality can be represented by the proposed embedding patterns.

To underpin the functioning of our differentiation between embedding \textit{and} representation patterns for network steganography, section \ref{Sect:Network_StegoRep} provides details for the integration of the known network hiding patterns into their corresponding \textit{representation} patterns.

\subsubsection{Example 2: Digital Media Steganography}
Commonly, three different approaches for generating steganographic digital media data exist, depending on the role of the underlying cover, which can be related to the Embedding Hiding Patterns proposed in Fig.~\ref{fig:taxonomy} as follows: Cover Modification modifies a pre-existing, non-steganographic carrier medium, for example by modulating the DCT coefficients in JPEG compressed images \cite{Fridich2009}, which can be categorized as EN5. In Cover Selection, the embedder generates subsets from a previously existing set of digital media, using specific attributes of the individual digital media to encode information. One possible method for this can be the choice of photographic images from a library.
Bits of value 0 or 1 are encoded by explicitly selecting portrait or landscape image orientations, respectively, and sequentially broadcasting them.
This falls into the category of EN2. Cover Synthesis describes the process of artificially generating digital media to embed the hidden data.
An example for pattern EN4 are computer-generated images, which can be composed in such way, that clip-arts are combined into an image and the actual selection from the clipart library builds the coding of the secret message (e.g., cars for a “1” , animals for a “0” message), rendering it a value that is modulated.

Continuous Digital Media (i.e., temporally changing media content like audio or video) further allow the modulation of temporal behavior as embedding pattern. For example, the inter-sample time intervals between samples of an audio stream can be artificially delayed or shortened in order to encode a hidden message, as an example for category ET1.1. 

\subsubsection{Example 3: Text Steganography}
Taking into account the taxonomy of the text hiding techniques from 
\cite{Ahvanooey:2019}, we can see how their taxonomy can naturally fit into our embedding hiding patterns. So we have:
\begin{itemize}
\item Structural methods – \textit{Open Space methods} involving the use of white or different Unicode spaces can be expressed with EN3 
or EN4 patterns. \textit{Line/Word shifting} which involves the position of a word in a line or of a line in a text, can be explained with EN2. \textit{Zero-Width methods} (by using ZWC Unicode characters that do not have text trace to represent different groups of $n$ secret bits) and \textit{Emoticons} use EN4. \textit{Feature/Format methods} can be explained with EN5.2 for characters and EN5 for other text elements (like paragraphs, sentences, etc).
\item Linguistic methods - \textit{Semantic methods} modifying the semantic attributes, such as spelling of words, abbreviations, synonyms, acronyms, paraphrasing, transliterations, and so on, can be expressed with EN4. \textit{Syntactic methods} which use changing of the diction and structure of text without significantly altering meaning or tone, such as ambiguous punctuation, shifting the location of the noun and verb, typographical errors, can be modeled via EN2 and EN1.
\item Random \& Statistics methods - \textit{Compression methods} (which hide the secret message in the compression codewords) and \textit{Random Cover methods} (which automatically generate cover message from some type, such as jokes, lists, notes, missing letter puzzles, Ci-poetry, etc., by using a secret bitstream, such as hiding in the first letter of the keyword) can be seen as hybrid methods. Both use a generated carrier from the secret bitstream, so this can be seen as a combination of Carrier \textbf{Size Feature Modulation} from EN5.1 and EN4.
\end{itemize}

\subsection{Representation Patterns}\label{Sect:NewTaxonomyRepresentation}

Here, essentially the same patterns can be applied as in the case of the embedding process. However, instead of describing how data is embedded, they describe how data is represented. Moreover, new patterns can be derived from representation patterns, which might not be directly reflected by embedding patterns.

As described in Sect.~\ref{EmbedNeqRepPattern}, representation patterns must not necessarily match embedding patterns during their application. Representation patterns can cover a larger variety of ideas than embedding patterns due to their domain-specific focus and because embedding patterns can \textit{cause} indirect actions, such as the termination of connections without actually \textit{performing} the termination. For instance, in network steganography, the patterns \textbf{Artificial (Forced) Reconnections Modulation} and \textbf{Artificial Retransmissions Modulation} have no direct counterpart at the side of embedding patterns. Such patterns are derived from their representation parent pattern (highlighted in \textbf{bold} font in Fig.~\ref{fig:taxonomy}).

In the remainder, we will cover network steganography representation patterns in detail to exemplify this concept. Providing a comprehensive taxonomy of representation patterns is part of our ongoing research.

\subsubsection{Network Steganography}\label{Sect:Network_StegoRep}

Fig.~\ref{fig:taxonomy} (right side) shows how network steganography representation patterns can be derived from embedding patterns.

Unfortunately, the current taxonomy of network steganography hiding patterns cannot be directly applied in the context of our novel taxonomy, as the distinction between timing and non-timing channels differs. The current taxonomy classifies \textit{more} patterns as temporal than our taxonomy (cf.\ Sect.~\ref{TemporalVSNontemporal}). Given that our definition of a temporal hiding pattern is a bit stricter than with the existing taxonomy, we would consider several \textit{protocol-aware} hiding patterns as non-temporal, in particular the original patterns:
\textbf{Artificial Loss},
\textbf{Artificial Reconnections}/\textbf{Retransmission} and
\textbf{Message Ordering} (determining which packet/connection is lost, retransmitted, reconnected, or in which order packets appear, is based on non-temporal attributes, such as TCP sequence numbers) as well as
\textbf{Temperature} (temperature is a state or value that can be modulated). 
However, the original pattern
\textbf{Frame Collisions} remains temporal and so do all protocol-agnostic timing patterns
(\textbf{Inter-packet Times}, \textbf{Message Timing}, and \textbf{Rate/Throughput}). However, their naming have been adjusted to the new terminology.

As also shown in Fig.~\ref{fig:taxonomy}, some representation patterns have no direct peer at the embedding patterns branch (bold derivations in the figure):

\begin{enumerate}
    \item \textbf{Frame Corruptions}: Frame collisions can be caused by timing a message (\textbf{Event Occurrence}) in a way that two messages collide. The hidden information is then represented by the collision. Also, \textbf{Frame Corruptions} is not a hybrid pattern as the content of the frame does not represent hidden information (and might be lost due to the collision), but the timing is the crucial information here for this pattern, i.e., \textit{when} a collision happens.
    \item \textbf{Artificial Retransmissions Modulation}: Several embedding actions can cause a retransmission, e.g., dropping selected TCP segments using the \textbf{Artificial Element-Loss Modulation} pattern or overloading a TCP buffer using the \textbf{Elements/Features Enumeration} pattern. However, the CR would observe the caused retransmission of PDUs.
    \item \textbf{Artificial (Forced) Reconnections Modulation}: same as in case of \textbf{Artificial Retransmissions Modulation}.
\end{enumerate}

Moreover, the following previous network steganography hiding patterns are now defined as \textbf{hybrid patterns}, which are not shown in Fig.~\ref{fig:taxonomy} to not burden the taxonomy:


\begin{enumerate}\label{hybrid_patterns}
    \item \textbf{Sequence Modulation}:\label{Ref:SequenceModulation} Because of its sub-patterns, this pattern modifies the \textit{position} of each element and their overall \textit{number}, which renders this pattern a hybrid form of \textbf{Elements/Features Enumeration} and \textbf{Elements/Features Positioning}.
    
    \item \textbf{Message Ordering} (former \textbf{PDU Ordering} pattern): This pattern orders PDUs instead of a message's elements. It is a sub-pattern of the \textbf{Sequence Modulation} hybrid pattern and now considered a non-temporal pattern (see Sect.~\ref{TemporalVSNontemporal}) as the \textit{position} of the PDUs and their \textit{number} are interpreted.

    \item \textbf{Add Redundancy} and \textbf{Modify Redundancy}: These patterns are known to do one of the following: 1) create additional space in a PDU to place hidden data (combination of \textbf{Size Feature Modulation} and \textbf{State/Value Modulation}); 2) compress data and then use the saved space to insert secret information, e.g., changing transmission codec for audio streams (transcoding steganography \cite{Wojciech:Transcoding}), which would mean that the \textbf{State/Value Modulation} pattern is applied twice in a row (first for compression and then the sub-pattern \textbf{Reserved/ Unused State/Value Modulation}).
\end{enumerate}

Taking advantage of our novel taxonomy, we were able to \textbf{discard} the following patterns from the network steganography domain as they \textit{mix embedding and representation} patterns.

\begin{enumerate}
    \item \textbf{Value Influencing} sub-pattern: This pattern was previously considered a sub-pattern of the original Value Modulation pattern. Some value is indirectly influenced by altering some surrounding condition that results in a modified value. However, this pattern actually groups two different patterns. The embedding pattern inserts hidden information by altering the surrounding value, but the representation pattern refers directly to the influenced value.
    
    \item \textbf{Payload Field Size Modulation}, \textbf{User-data Value Modulation \& Reserved Unused}: Their concepts are already found in the respective patterns \textbf{Size Feature Modulation}, \textbf{State/Value Modulation} and \textbf{Reserved/Unused State/ Value Modulation}, from which they were derived. As discussed in Sect.~\ref{Sect:UserdataPayload}, we discarded the distinction between payload and non-payload.
    
    \item \textbf{User-data Corruption} pattern: This pattern refers to hybrid methods such as HICCUPS or RSTEG, which, e.g., retransmit a message and then replace the original content. The overwriting however must be considered as \textbf{Blind State/Value Modulation} whereas the retransmission refers to the new \textbf{Artificial Retransmissions} pattern. While this would render the pattern a hybrid one, it was discarded due to the the same reason as \textbf{Payload Field Size Modulation} was.
    
    \item \textbf{Temperature} pattern: Similarly to the \textbf{Value Influencing} pattern, embedding and representation must be split. The secret data is represented by some temperature value but is embedded by, e.g., high CPU load. Moreover, this pattern is domain overlapping: network load can influence the CPU temperature (network-specific pattern) while the temperature value is a physical value, belonging to the domain of CPS steganography.
\end{enumerate}

\subsubsection{Other Steganography Domains}

As discussed, this paper focuses on embedding patterns. Thus, representation patterns were only illustrated for the network steganography domain. Future work will extend the taxonomy to cover representation patterns for additional domains, especially digital media, text and cyber-physical systems steganography.

\section{Anticipated Steganography Developments in the Context of Patterns}\label{Sect:AnticipatedDevs}

Our proposed pattern-based taxonomy needs to proof its functionality under the umbrella of future trends, such as:

\begin{enumerate}
     \item \textbf{Novel Application Domains for Steganography.} We expect several new domains of steganography to emerge during the next decade. As pointed out by Bezahaf et al., the Internet will be required to adapt to certain requirements of new services, including holographic applications, autonomous vehicles, remote surgery, and automated reality \cite{InternetEvolution20}. Such services will provide several new options for the embedding of digital media steganography and CPS steganography. These new services will exploit 5G+ and low-earth-orbit satellite clusters while being linked to higher performance characteristics in terms of Quality of Service \cite{InternetEvolution20}, which will provide novel communication protocols that will allow enhanced forms of network steganography. It cannot be stated whether novel hiding patterns will emerge during these developments but their application scenarios will widen.
    
    \item \textbf{Steganography for Machine Learning (ML) Systems.}
    Attackers could exploit ML systems and its related processes to embed secret information. The recent body of research has shown that ML can be influenced by adversary attacks, overfitting (eases manipulations) and data poisoning, among other aspects \cite{MLSec:Top10}. 
    This development is reflected in the ongoing work to establish a taxonomy for such attacks in form of the so-called Adversarial ML Threat Matrix by the MITRE Corporation and others \cite{spring_adversarial_2020}. 
    For instance, adversary manipulations of road signs for smart vehicles can lead to false categorizations of such signs. However, the process from data-collection to generation of ML-based outputs can be potentially influenced by a steganographer. For instance, one could try to influence certain aspects of 
    raw data in a way that the ML system might provide excellent outputs in practice. However, when minimal changes to specific parts of the input data are conducted, results might differ. The output of an ML system could then represent a secret message and the modification of input data would be the steganographic key. Alternatively, secret data could be nested directly inside the ML models. A first paper that exploits federated learning for steganography is \cite{costa2021covert}. Again, ML steganography might lead to novel hiding patterns. 

    \item \textbf{Adaptive Countermeasures.} Current steganography countermeasures are usually tailored for testbed environments, where they provide sufficient results, see \cite{caviglione2021trends} for a comprehensive overview. However, not only do real-world applications demand very low false-positive rates as false-positives accumulate to large numbers for large-scale scenarios \cite{steinebach2018channelsteganalysis}, they also have to deal with continuously changing data. For instance, since the invention of the ARPANET, the Internet's traffic characteristics continuously kept changing \cite{InternetEvolution20}. This is a significant problem since many detection methods are tailored for Internet or network traffic provided at a given time and under specific environmental attributes. For this reason, countermeasures need to be adaptive. First approaches were already tailored, such as the \textit{dynamic warden} \cite{FGCS:DynWarden}. Further research paths for countermeasures might exploit multi-agent systems (MAS) that simulate (large-scale) network environments attacked by steganography, thus allowing predictable behaviour of stego-malware and adjustment of countermeasures. At the moment, it is still unclear, how dynamic countermeasures are linked to the characteristics of specific patterns.
    
    \item \textbf{Hybrid Transfer and Storage as well as Chaining of Patterns.} Our new taxonomy allows describing hiding techniques in a much more precise manner than before. Consider, for instance, the recently introduced DeadDrops \cite{Schmidbauer:DeadDrops:ARP} which exploit network protocol caches for storage while they secretly transfer the information over a network covert channel to embed the secret data inside the caches. Such methods apply one hiding pattern for embedding of secret information into the transfer from the CS to the DeadDrop, further embed the information using a second pattern (e.g., to alter an NTP or ARP cache), while the CR might indirectly retrieve the information using some third representation pattern. However, such a \textit{chaining} of embedding patterns in a way that multiple steganography domains are utilized is not well-understood yet.

\end{enumerate}

\section{Conclusion and Future Work}\label{Sect:ConclFut}

We revised the entire taxonomy of hiding patterns. Our new taxonomy provides a tool for \textit{all} domains of steganography -- not solely network steganography -- and thus allows the utilization of hiding patterns also in digital media, text, CPS, filesystem, and other steganography areas. We also provide a clearer distinction between the embedding process and the representation by hidden patterns than available through the current taxonomy. Wherever suitable, we kept previous terms in order to maximize backward-compatibility with the old taxonomy and to ease the transition for users of the previous taxonomy.

The next steps of our consortium will be to address the following topics in order to develop our proposed taxonomy further:
We plan to extend the size of our consortium so that more stakeholders from additional domains, such as filesystem steganography and CPS steganography, can contribute to it. We will further extend the representation patterns taxonomies to fully reflect each domain. This will aid the further distribution and acceptance of the model while also improving its functionalities and widening its application domain. 
Moreover, we currently evaluate the integration of linked domains, such as digital watermarking, into the taxonomy, which would require the inclusion of additional experts into our consortium.

\textbf{Acknowledgements}
Parts of the work from Brandenburg and Magdeburg authors in this paper (i.e., on definitions and general discussions) have been funded by the German Federal Ministry for Economic Affairs and Energy (BMWi, Stealth-Szenarien, Grant No. 1501589A and 1501589C) within the scope of the German Reactor-Safety-Research-Program. 

Parts of the work of Laura Hartmann has been funded by the European Union from the European Regional Development Fund (EFRE) and the State of Rhine-land-Palatinate (MWWK), Germany. Funding content: P1-SZ2-7 F\&E: Wissens- und Technologietransfer (WTT), Application number: 84003751, project MADISA. Her work has also been funded by Programm zur Förderung des Forschungspersonals, Infrastruktur und forschendem Lernen (ProFIL) of the University of Applied Sciences Worms.

Parts of the work of Luca Caviglione and Wojciech Mazurczyk have been supported by the SIMARGL Project - Secure Intelligent Methods for Advanced RecoGnition of malware and stegomalware, with the support of the European Commission and the Horizon 2020 Program, under Grant Agreement No. 833042.

\bibliographystyle{abbrv}
\bibliography{main}

\end{document}